\documentclass[12pt]{article}

\usepackage{dcolumn}
\usepackage{bm}
\usepackage[latin1]{inputenc}
\usepackage[spanish,english]{babel}
\usepackage{amsfonts}
\usepackage{amssymb}
\usepackage{graphicx}
 \usepackage{setspace}

\newcommand{\be}{\begin{equation}}
\newcommand{\ee}{\end{equation}}
\newcommand{\bea}{\begin{eqnarray}}
\newcommand{\eea}{\end{eqnarray}}

\begin{document}

\begin{titlepage}

\begin{center}
{\large \bf Inflation, quantum fields, and CMB anisotropies\footnote{Essay awarded with the fourth prize in the Gravity Research Foundation 2009 Essay Competition on Gravitation} } \\
\end{center}

\begin{center}
Iván Agulló\footnote{ ivan.agullo@uv.es}\\ 
\footnotesize \noindent {\it Physics Department, University of
Wisconsin-Milwaukee, P.O. Box 413, Milwaukee, WI 53201, USA}\\
and\\
\footnotesize \noindent {\it Departamento de Física Teórica and
IFIC,  Universidad de Valencia-CSIC, Facultad de Física,
Burjassot-46100, Valencia, Spain.}
\end{center}
\begin{center}
José Navarro-Salas\footnote{ jnavarro@ific.uv.es}\\

\footnotesize \noindent {\it Departamento de Física Teórica and
IFIC,  Universidad de Valencia-CSIC, Facultad de Física,
Burjassot-46100, Valencia, Spain.}
\end{center}

\begin{center}
Gonzalo J. Olmo\footnote{ olmo@iem.cfmac.csic.es}

\footnotesize \noindent {\it Perimeter Institute for Theoretical
Physics, Waterloo, Ontario, N2L 2Y5 Canada and\\ Instituto de
Estructura de la Materia, CSIC, Serrano 121, 28006 Madrid, Spain}
\end{center}

\begin{center} Leonard Parker\footnote{ leonard@uwm.edu}\\

\footnotesize \noindent {\it Physics Department, University of
Wisconsin-Milwaukee, P.O. Box 413, Milwaukee, WI 53201, USA}

\end{center}
\bigskip
\begin{center}
{\it March 4, 2009}
\end{center}

\newpage
\begin{abstract}

Inflationary cosmology has proved
to be the most successful at predicting the properties of the anisotropies
observed in the cosmic microwave background (CMB). In this essay we
show that  quantum field renormalization significantly
influences the generation of primordial perturbations and hence the expected measurable imprint of cosmological inflation
on the CMB. However, the new
predictions remain in agreement with observation, and in fact  favor the
simplest forms of inflation.
In the near future, observations of the influence
of gravitational waves from the early universe on the CMB will test  our
new predictions.

\end{abstract}

\end{titlepage}

One of the most exciting ideas of contemporary physics is to
explain the origin of the observed structures in our universe as a
result of  quantum fluctuations in the  early expanding  universe. As first
shown in the sixties \cite{parker69}, the amplification of quantum
field fluctuations is an unavoidable consequence in a strongly time-dependent gravitational field \cite{parker-toms,
birrel-davies}. Fundamental physical implications were implemented some years
latter to culminate, in the seventies,  with the prediction of the evaporation of black holes with a black-body spectrum
\cite{hawking} and, in the eighties, when the inflationary model of the universe was
 introduced \cite{inflation}, predicting that
small density perturbations are likely to be generated in
 the very early  universe  with a nearly scale-free spectrum
\cite{inflation2}. In the nineties, the detection of
temperature fluctuations in the cosmic microwave background
(CMB) by the COBE satellite \cite{cobe}  appeared to be
consistent with the inflationary cosmology predictions. In
the present decade,  the predictions of inflation have been
confirmed in the specific pattern of anisotropies imprinted
in the full sky map of the CMB, as reported, for instance,
by the WMAP mission \cite{WMAP5}.
Moreover,  an inflationary-type expansion also
predicts the creation of primordial gravitational waves \cite{grishchuk-starobinsky}, whose effects still
remain undetectable.
Forthcoming experiments, such as the PLANCK satellite \cite{planck},  may measure effects of relic gravitational
waves and offer new trends for gravitational physics in the next decade.
Therefore, it is particularly important to scrutinize the quantitative predictions of quantum field theory  in an inflationary background. This is the aim of this essay.

As remarked above, a strongly time-dependent gravitational field necessarily amplifies vacuum fluctuations. This happens, typically, in a rapidly expanding universe and also in a gravitational collapse. The event horizon of a black hole acts as a magnifying glass that exponentially stretches very short wavelengths to macroscopic scales and generates a thermal flux of outgoing particles. Similarly, during  exponential inflation, $ds^2= -dt^2 + e^{2Ht}d\vec{x}^2$, a typical physical length, with comoving wavenumber $k$, increases exponentially $k^{-1}e^{Ht}$  and reaches the Hubble radius, $H^{-1}=$constant, at some time  $t_k$ ($ke^{-Ht_k}=H$). These quantum fluctuations produce scale-free density perturbations and relic gravitational waves via a quantum-to-classical transition at the time  of Hubble horizon exit $t_k$.  The cosmic expansion farther stretches these scale-free primordial perturbations to astronomical scales.

Let us focus on the production of relic gravitational waves by considering fluctuating tensorial modes $h_{ij}(\vec{x},t)$ in an exponentially expanding, spatially flat universe: $g_{ij}=a^2(t)(\delta_{ij} + h_{ij})$, with $a(t)= e^{Ht}$. The
perturbation field $h_{ij}$ can be decomposed into two polarization
states described by a couple of  massless scalar fields $h_{+,
\times}(\vec{x},t)$, both obeying the wave equation  $ \label{waveq}
\ddot{h} + 3H \dot{h} -a^{-2}\nabla^2 h
=0$ (see, for instance, \cite{books}; we omit the
subindex $+$ or $\times$).
On scales smaller than the Hubble radius the spatial gradient term dominates the damping term  and leads to the conventional
flat-space oscillatory behavior of modes. However, on scales larger than the Hubble radius the damping term $3H \dot{h}$ dominates. If one considers
 plane wave modes of comoving wavevector $\vec{k}$ obeying the adiabatic condition \cite{parker69,parker-toms,
birrel-davies}
and de Sitter invariance one finds \be \label{modesh}
h_{\vec{k}}(\vec{x},t) = \sqrt{\frac{16\pi G}{2
(2\pi)^3k^3}}e^{i\vec{k}\vec{x}} ( H
-ike^{-Ht})e^{i(kH^{-1}e^{-Ht})} \ . \ee
 These modes oscillate until the physical wave length reaches the Hubble horizon length. A few Hubble times after
 horizon exit the modes amplitude freezes out to the constant value $|h_{\vec{k}}|^2 = \frac{ GH^2}{\pi^2k^3}$.  Because of the loss of phase information, the modes of the perturbations
soon take on classical properties
\cite{KieferPointerStates2007}. The freezing amplitude is
usually codified through the quantity $\Delta_h^2(k)\equiv 4\pi k^3 |h_{\vec{k}}|^2$.
Taking
into account the two polarizations,   one easily gets the standard
scale free tensorial power spectrum \cite{books} $P_t(k) \equiv 4\Delta_h^2(k)=
\frac{8}{M_P^2} \left (\frac{H}{2\pi}\right)^2$, where $M_P=1/\sqrt{8\pi G} $.
It is easy to see that $\Delta_h^2(k)$ gives the formal contribution, per $d\ln k$, to the variance of the gravity wave fields $h_{+,
\times}$
 \be \label{varianceh}
\langle h^2 \rangle =
\int_0^{\infty}\frac{dk}{k}\Delta^2_h (k) \ .
\ee
Due to the large $k$ behavior of the modes the above integral is divergent. It is a common view \cite{books} to
bypass this point by regarding $h(\vec{x}, t)$ as a classical random field. One then introduces a window function $W(kR)$, multiplying at $\Delta^2_h (k)$ in the integral, to smooth out the field on a certain scale $R$ and to remove the Fourier modes with $k^{-1} < R$.
One can also consider unimportant the value of $\langle h^2 \rangle$ and regard the (finite) two-point function $\langle h(x_1)h(x_2) \rangle$, uniquely defined by $\Delta^2_h (k)$, as the basic object.
However, $\langle h^2 \rangle$ represents the
variance of the Gaussian probability distribution associated to
$h(\vec{x}, t)$, which means that at any point $h(\vec{x}, t)$ may
fluctuate by the amount $\pm\sqrt{ \langle h^2(\vec{x}, t)
\rangle}$ defining this way a classical perturbation.
 It is our view to regard the variance as the basic physical object and treat $h$ as a proper quantum field. Renormalization is then the natural solution to keep the variance finite and well-defined.
Since the physically relevant quantity (power spectrum) is expressed
in momentum space, the natural renormalization scheme to apply is
the so-called adiabatic subtraction \cite{parker07}, as it
renormalizes the theory in momentum space. Adiabatic renormalization
\cite{parker-fulling, parker-toms, birrel-davies} removes the
divergences present in the formal expression (\ref{varianceh}) by
subtracting counterterms 
mode by mode in the integral (\ref{varianceh}) \be
\label{variancehren} \langle h^2 \rangle_{ren} =
\int_0^{\infty}\frac{dk}{k}[4\pi k^3|h_{\vec{k}}|^2 - \frac{16\pi G
k^3}{4\pi^2 a^3}(\frac{1}{w_{k}}+\frac{\dot{a}^2}{2a^2w_{k}^3}
+\frac{\ddot{a}}{2aw_{k}^3})]\ ,\ee with $w_k=k/a(t)$. The
subtraction of the first term $(16\pi G k^3/4\pi^2 a^3 w_k)$
cancels the typical flat space vacuum fluctuations.
However, the additional terms, proportional to
$\dot{a}^2$ and $\ddot{a}$ are necessary to properly perform the
renormalization in an expanding universe.

For the idealized case of a strictly constant $H$, the
subtractions exactly cancel out the vacuum amplitude
\cite{parker07}, at any time during inflation, producing a
vanishing result for  the variance. Therefore, the physical
tensorial power spectrum, the integrand of (\ref{variancehren}),
is zero. Note that this surprising result does not contradict the
fact that quantum fluctuations in de Sitter space produce a
Hawking-type radiation with temperature $T_H=H/2\pi$
\cite{gibbons-hawking}. This temperature stems from the comparison
of the modes (\ref{modesh}) with those defining the vacuum of a
static observer, located at the origin of coordinates, with metric
$ds^2=-(1-H^2r^2)d\tilde{t}^2+ (1-H^2r^2)^{-1}dr^2 +
r^2d\Omega^2$. The different time/phase behavior of both sets of
modes (captured by non-vanishing Bogolubov coefficients) produces
the Hawking temperature. However, their amplitudes are  exactly
the same \cite{agullo-navarro-olmo-parker08}.

Does it mean that inflation does not produce gravitational waves?
No. For more realistic inflationary models, $H\equiv \sqrt{\frac{8\pi
G}{3}V(\phi_0)}$ slowly decreases  as $\phi_0$ (the classical homogeneous part of
the inflaton field) rolls down the potential towards a minimum.
Tensorial perturbations are then expected, on dimensional grounds, to be produced with amplitude
proportional to $\dot{H}$, instead of $H^2$. The form of the modes
is now $h_{\vec{k}}(t, \vec{x}) = (-16\pi G\tau
\pi/4(2\pi)^3a^2)^{1/2}$\ $\times H^{(1)}_{\nu}(-k\tau)e^{i\vec{k}\vec{x}}$,
were the index of the Bessel function is  $\nu=\sqrt{9/4
+3\epsilon}$ and  $\epsilon$ is the slow-roll parameter
$\epsilon\equiv -\dot{H}/H^2= (M^2_P/2)(V'/V)^2$. The conformal
time $\tau\equiv\int dt/a(t)$ is given here by
$\tau=-(1+\epsilon)/aH$. The loss of phase information in the modes
still occurs at a few Hubble times after horizon exit,
converting the fluctuations to classical perturbations. Therefore, it is natural to evaluate the new integrand of (\ref{variancehren}) (i.e, the tensorial power
spectrum) a few
Hubble times after the time $t_k$. Since the results will not be
far different from those at $t_k$, we use the time $t_k$ to
characterize the results. The new tensorial power spectrum  turns
out to be then \be \label{psT2}P_t(k) = \frac{8\alpha }{M_P^2}
\left(\frac{H(t_k)}{2\pi}\right )^2 \epsilon(t_k) \equiv
-\frac{8\alpha }{M_P^2} \dot{H}(t_k) \ ,
 \ee
where $\alpha\approx 0.904$ is a numerical coefficient.
As expected,
it is just the deviation from an exactly constant $H$, parameterized by
$\epsilon$, which generates a non-zero tensorial power spectrum.

The above result would then imply that the tensor to scalar ratio $r= P_t(k)/P_{\cal R}(k)$ may be well below the standard predictions of single-field inflationary models. However, this is not necessarily the case since the scalar power spectrum $P_{\cal R}(k)$, which constitutes the seeds for structure formation, is also affected by renormalization. A detailed calculation, sketched in \cite{agullo-navarro-olmo-parker09}, leads to
\be \label{sps2}P_{\cal R}=
\frac{1}{2M_p^2\epsilon(t_k)}\left(\frac{H(t_k)}{2\pi}\right)^2(\alpha\epsilon(t_k)
+3\beta{\eta}(t_k)) \ , \ee where $\beta \approx 0.448$ is
numerical coefficient and $\eta \equiv
M_P^2(V''/V)$ is  the second slow-roll parameter.
Note that this contrasts with  the standard prediction: $P_{\cal R}=
\frac{1}{2M_p^2\epsilon(t_k)}\left(\frac{H(t_k)}{2\pi}\right)^2$ \cite{books}. Since the scalar amplitude is also modulated by the slow-roll parameters the ratio $r$ is given by
\be r= \frac{16\epsilon^2(t_k)\alpha}{\alpha\epsilon(t_k)
+3\beta{\eta}(t_k)} \ , \ee which contrasts with the standard result $r= 16\epsilon (t_k)$. To translate this difference to a closer empirical level
 we have to
introduce  the scalar and tensorial spectral indices $n_s\equiv 1+ d
\ln P_{\cal R}/d\ln k, \ n_t \equiv d \ln P_{t}/d\ln k$, and the running tensorial index $n'_t\equiv
d{n_t}/d\ln{k}$.  The
standard expression for  the relation between the
tensor-to-scalar ratio $r\equiv P_t/P_{\cal {R}}$ and spectral indices (consistency condition) is: $r= -8n_t$. It is expected to be verified by any single-field slow-roll inflationary model, irrespective of the particular form of the potential.
However, if we
invoke renormalization we get a more involved consistency condition $r=r(n_t, n_s, n'_t)$.  For illustrative purposes, in the
simplest case of $n'_t\approx 0$ and taking the approximation $\alpha
\approx 2 \beta$, the new consistency condition becomes \be
r=1-n_s+\frac{96}{25}n_t+\frac{11}{5}\sqrt{(1-n_s)^2+\frac{96}{25}
n_t^2}\ . \ee
Note that this expression allows for a null tensorial tilt  $n_t\approx 0$ while being compatible with a non-zero ratio $r\approx \frac{16}{5}(1-n_s)$.

We can  compare the new predictions with
the standard ones on the basis of the five year WMAP results.   We find \cite{agullo-navarro-olmo-parker09}, see Figure 1,  that the new
predictions agree with observation and improve the likelihood that the
simplest potential energy functions (quadratic and quartic, respectively)
are responsible for driving the early inflationary expansion of the universe. The influence of relic gravitational waves on the CMB will soon
come within the range of  planned satellite measurements, and this will be a definitive test of the new predictions.
\begin{figure}[htbp]
\begin{center}
\includegraphics[width=6.6cm]{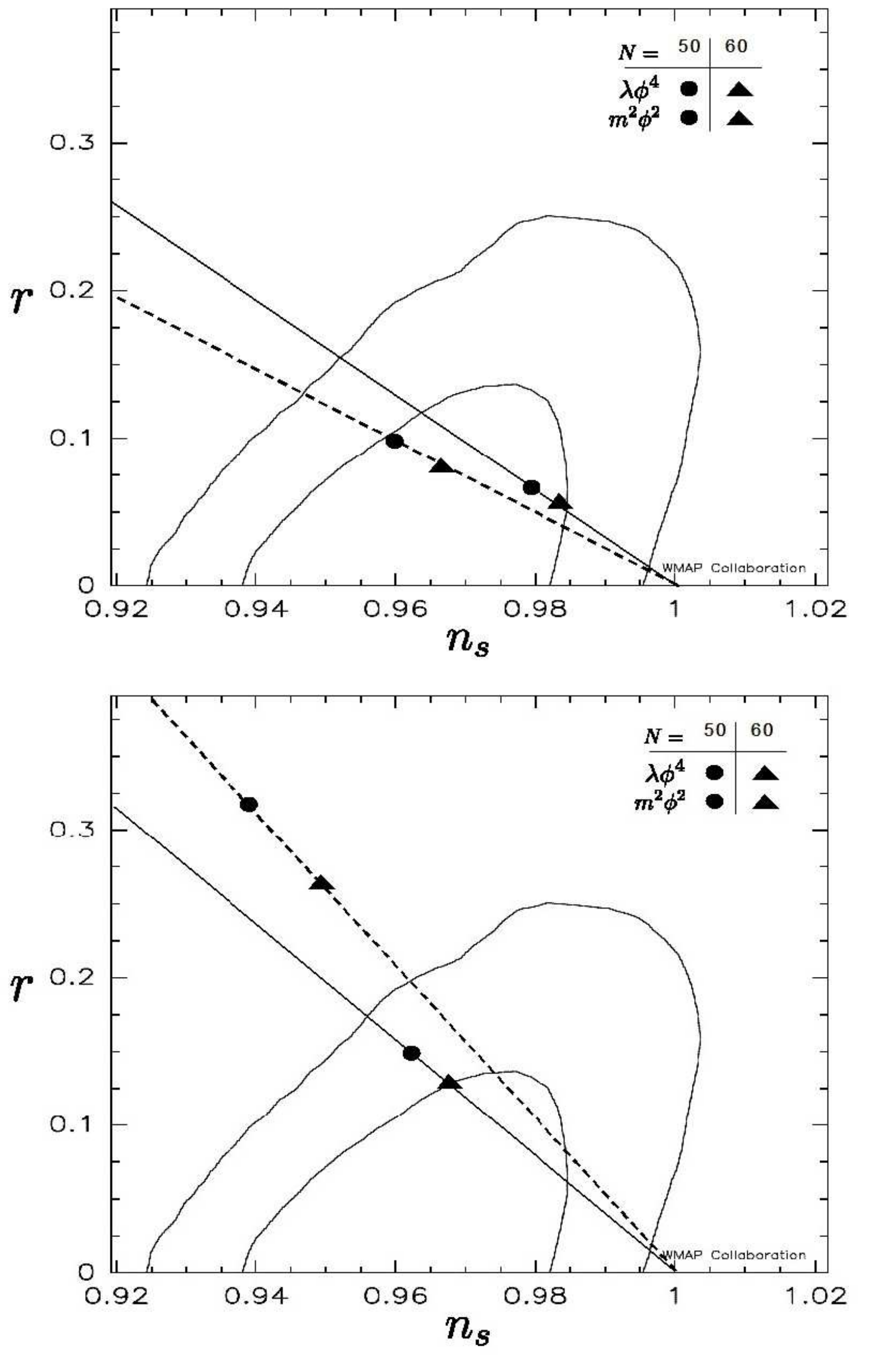}
\caption{Plot of $r$ versus $n_s$. The contours show the $68\%$
and $95\%$ CL derived from WMAP5 (in combination with BAO$+$SN)
\cite{WMAP5}. We consider two representative inflation models:
$V(\phi)= m^2 \phi^2$ (solid line), $V(\phi)= \lambda \phi^4$
(dashed line). The symbols show the prediction from each of this
models in terms of the number $N$ of e-folds of inflation for the
monomial potentials. The top part corresponds to the prediction of
our formulae, while the bottom one corresponds to the standard
prediction.}
\end{center}
\end{figure}



\end{document}